\documentclass{article}
\usepackage{slashed,feynmf,epsfig,amsmath,amssymb,enumitem}
\usepackage[papersize={8.5in,11in}]{geometry}
   \usepackage{bm}
\usepackage{graphicx}
\newcommand{\be}{\begin{equation}}
\newcommand{\ee}{\end{equation}}
\begin{document}
\title{Ultraviolet Complete Quantum Field Theory and Particle Model}
\author{J. W. Moffat\\
Perimeter Institute for Theoretical Physics, Waterloo, Ontario N2L 2Y5, Canada}
%and\\
%Department of Physics and Astronomy, University of Waterloo, Waterloo,\\
%Ontario N2L 3G1, Canada}
\maketitle

%\thanks{PACS:}

\begin{abstract}
An ultraviolet complete particle model is constructed for the observed particles of the standard model. The quantum field theory associates infinite derivative entire functions with propagators and vertices, which make quantum loops finite and maintain Poincar\'e invariance and unitarity of the model. The electroweak model $SU(2)\times U(1)$ group is treated as a broken symmetry group with non-vanishing experimentally determined boson and fermion masses. A spontaneous symmetry breaking of the vacuum by a scalar Higgs field is not invoked to restore boson and fermion masses to the initially massless $SU(2)\times U(1)$ Lagrangian of the standard model. The hierarchy naturalness problem of the Higgs boson mass is resolved and the model contains only experimentally observed masses and coupling constants. The model can predict a stable vacuum evolution. Experimental tests to distinguish the standard model from the alternative model are proposed. The finite quantum field theory can be extended to quantum gravity.
\end{abstract}

\maketitle

%\date{\today}

\section{Introduction}

We will address the following question: If at the beginning of the development of the standard model (SM), quantum field theory (QFT) had been a finite theory free of all divergencies, how would the SM have been formulated? The 12 quarks and leptons, the weak interaction $W^\pm$, $Z^0$ vector bosons and the scalar Higgs boson would be discovered and complete the particle content of the SM. However, guaranteeing the renormalizability of the electroweak (EW) sector would not have been a problem needing resolution, because a perturbatively finite QFT would be the foundation of the theory. The motivation of the SM is through symmetry. The renormalizability of the model requires a gauge invariance symmetry. This succeeds for QCD and QED, for the eight colored gluons and the photon are massless guaranteeing gauge invariance. To ensure renormalizability of the electroweak sector $SU(2)\times U(1)$, it is assumed as an initial postulate that all boson and fermion masses are zero leading to an $SU(2)\times U(1)$ invariant Lagrangian. The particle masses are restored in the EW sector by a spontaneous Higgs mechanism, invoked through the addition of a scalar degree of freedom and a broken symmetry vacuum maintaining the $SU(2)\times U(1)$ invariance of the Lagrangian.

In an earlier paper~\cite{Moffat1990}, a finite QFT formulation including a Higgs boson field was developed with an initially gauge invariant symmetric $SU(2)\times U(1)$ with zero $W$ and $Z$ boson and fermion masses and the gauge symmetry was broken by a spontaneously broken vacuum~\cite{Higgs1,Higgs2,BroutEnglert,Kibble}. In the following, we will derive an alternative model with the underlying QFT of the particle model finite to all orders of perturbation theory, and recognizing that $SU(2)\times U(1)$ is a {\it badly broken symmetry} due to the masses of the particles and not demanding spontaneous symmetry breaking of the vacuum. A finite QFT leads naturally to a physical energy scale through the length $\ell_M=1/\Lambda_M$, retaining Poincar\'e invariance, unitarity and analyticity of scattering amplitudes. We will demonstrate how the finite QFT particle model can solve the Higgs mass fine-tuning and naturalness problem, while agreeing with all CERN accelerator experiments.

The SM of particle physics is successful when compared to experimental data~\cite{Dawson}. The discovery of the Higgs boson with a mass $m_H= 125\,{\rm GeV}$ supported the standard scenario of the spontaneous symmetry breaking of the EW group $SU(2)\times U(1)$ through a non-zero vacuum expectation value of the complex Higgs scalar field. The determination of the Higgs boson mass in renormalization theory leads to a severe fine-tuning and naturalness problem~\cite{Wells}. The field potential has the Ginsburg-Landau form~\cite{Landau}:
\be
\label{GinsbergLandaupotential}
V(\phi)=-\mu^2\phi^\dagger\phi+\lambda(\phi^\dagger\phi)^2.
\ee
The Higgs mass in the model is determined to be
\be
m_H^2=m_{\rm bare}^2+\delta m_H^2,
\ee
where $m_H$ is the Higgs boson mass, $m_{\rm bare}$ is the Higgs boson bare mass of the unrenormalized Lagrangian, and the radiative correction $\delta m_H^2$ is given by
\be
\delta m_H^2=\frac{y_t^2}{16\pi^2}\Lambda_C^2+\delta{\cal O}(m_{\rm weak}^2).
\ee
Here, $y_t$ is the top quark Yukawa coupling of order 1, and $\Lambda_C$ is the energy cut-off to the quadratically divergent Higgs-top quark loop momentum integral of the Higgs boson self-energy, and $\delta{\cal O}(m_{\rm weak}^2)$ are other quantum corrections where $m_{\rm weak}$ denotes the weak scale. There are many more loop correction contributions, in addition to the top quark. The fine-tuning of $m_H^2$ is avoided, if $\mu^2$ is adjusted, $\delta m_H^2\sim  m_H^2$ and $\Lambda_C\lesssim 1\,{\rm TeV}$. However, the standard model is expected to be valid up to the Planck energy, $\Lambda_C\sim 10^{19}\,{\rm GeV}$, and any new particles above $1\,{\rm TeV}$, such as a see-saw neutrino at an energy $E\sim 10^{13}\,{\rm GeV}$ range would create a severe naturalness fine-tuning problem.

In the SM the absolute masses of the $W^\pm$ and $Z^0$ bosons and the quarks and leptons are not determined by calculation. The masses of fermions determined by the Yukawa Lagrangian and the EW spontaneous symmetry breaking scale $v=246\,{\rm GeV}$ are not predicted by the model. The coupling constants needed to determine the masses $m_f=g_fv$ are arbitrary and are added in an {\it ad hoc} way as free parameters corresponding to the experimentally known masses of the standard model. The mass of the Higgs boson fails to be determined because of the quadratic divergence of the radiative Higgs loop graphs.

The Higgs mass fine-tuning problem could be eliminated if supersymmetric partners were discovered, or new particles such as top quark partners were discovered that can cancel the quadratic divergences of the loop integrals in the radiative correction $\delta m_H^2$. The LHC at CERN has not discovered supersymmetric particles or new particles, which can eliminate the Higgs mass hierarchy fine-tuning problem.

The SM is based on the assumption that the $W$ and $Z$ boson masses and the quark and lepton masses are initially zero allowing for the gauge invariance symmetry of $SU(2)\times U(1)$~\cite{Weinberg,Salam,thooft,thooftVeltman,Taylor,Halzen,Aitchison,Peskin}. The {\it ad hoc} choice of the scalar field potential (\ref{GinsbergLandaupotential}) and an imaginary mass spontaneously breaks the symmetry of the vacuum and three Goldstone bosons, which are absorbed by the gauge bosons $W^\pm$ and the $Z^0$, leaving the $U(1)$ photon massless.  The ``EW unification" is not as complete as the unification of the electric and magnetic fields in Maxwell's electromagnetic theory, for the weak coupling constants $g$ and $g'$ are separate constants.

In the following, we present a model which can eliminate the Higgs mass fine-tuning and naturalness problem and stabilize the vacuum evolution in the Universe. It can also incorporate massive neutrinos. The model accepts from the beginning that $SU(2)\times U(1)$ {\it is a broken symmetry group} with non-zero experimentally measured $W$, $Z$, Higgs boson masses and quark and lepton masses. The problem of infinite renormalizability is resolved by regulated propagators and vertices for the Feynman loop diagrams of QED, QCD and weak interactions.  The propagators are regulated by an infinite derivative entire function ${\cal E}(p^2)$, which is  analytic and holomorphic in the complex $p^2$ plane with a pole and/or an essential singularity at $p^2\rightarrow\infty$. The QFT forming the basis of the model is finite, unitary and Poncar\'e invariant to all orders of perturbation theory~\cite{Moffat1990,MoffatWoodard1991,Moffat1991,Moffat2011,Moffat2011(2),KleppeWoodard1993}. A derivation of a finite QFT based on higher spin fields has been developed~\cite{Moffat1989}. In previous work, the finite QFT was applied to the SM with a spontaneous symmetry breaking mechanism and an initial $SU(2)\times U(1)$ gauge symmetry~\cite{Moffat1990}. The finite QFT was also extended to quantum gravity and a solution to the cosmological constant problem~\cite{Moffat1990,MoffatWoodard1991,Moffat2011(2),Moffat2014}. In the following, we will not invoke a spontaneous symmetry breaking of the vacuum, treating the EW group $SU(2)\times U(1)$ as a broken group due to the experimentally determined masses of the bosons and fermions.

We will relax the QFT strong assumption of polynomial behavior of amplitudes at infinity. This development employs the introduction of {\it entire functions} in momentum space, which preserves unitarity, for no additional unphysical singularities are introduced at finite energies. The amplitudes have poles or an essential singularity at infinity. The presence of an essential singularity at infinity can destroy the process of going from Minkowski space to Euclidean space by rotating the contour of integration over the energy ($p_0\rightarrow ip_4)$. However, regularized entire functions can be constructed that allow the QFT to be formulated from the outset in Euclidean momentum space, and then allow an analytic continuation to Minkowski space~\cite{Moffat2011}. In finite QFT there is no fundamental difference between renormalizable and non-renormalizable theories. The loop radiative corrections in finite QFT decrease rapidly enough in Euclidean momentum space, guaranteeing their finiteness. A violation of perturbative unitarity of scattering amplitudes is avoided by the exchange of the scalar Higgs boson between the SM particles. The Higgs mass hierarchy problem is resolved by having the Higgs coupling $\Lambda_H$ energy scale controlling Higgs-fermion and Higgs self-interaction loops satisfy, $\Lambda_H\lesssim  1\,{\rm TeV}$. This energy condition can be checked by future experiments determining the strength of the radiative couplings of the Higgs particle to the SM particles in loop diagrams. The other particle loop diagrams will be controlled by an energy scale $\Lambda_M > 1\,{\rm TeV}$.

If there is a single mass scale $M$ in QFT theory, then an effective field theory can be implemented as an expansion in $1/M$~\cite{Peskin,Burgess}. The construction of the effective field theory accurate to some power of 1/M requires a new set of free parameters at each order of the expansion in 1/M. Since effective field theories are not valid at small length scales, they need not be renormalizable. Indeed, the ever expanding number of parameters at each order in 1/M required for an effective field theory means that they are generally not renormalizable in the same sense as QED, which requires only the renormalization of two parameters, namely, the mass and charge of the electron. In practice, an effective QFT ignores all the particles too heavy to be produced. Application of the renormalization group method modifies the high energy behavior of QFT, so that the effective theory is only a valid description of the physics at energies below the masses of heavy particles. In the SM the increase in the number of free parameters can grow significantly with increasing inverse powers of $1/M$, curtailing the predictive power of the effective field theory. In contrast, a finite QFT, if correct, is valid to all energies and can predict particle interactions and scattering amplitudes to any order in perturbation theory with a limited number of parameters. Moreover, in our particle model, we claim that there are no new particles to be observed beyond the discovered 12 bosons and 12 quarks and leptons, so this precludes the use of an effective field theory with new massive particles.

The finite QFT model is based on $SU(3)\times SU(2)\times U(1)$ and QCD is formulated for a finite Yang-Mills, non-Abelian QFT and symmetric color group $SU(3)_C$. The eight color gluons are massless and the symmetric gauge invariant QFT possesses an energy dependent vertex coupling ${\bar g}_s(p^2)=g_S{\cal E}(p^2/\Lambda_S^2)$ where $\alpha_S=g_S^2/4\pi$ is the strong interaction coupling constant. Gauge invariant QED with a massless photon will possess an energy dependent coupling constant ${\bar e}(p^2)=e{\cal E}(p^2)$, controlled by the energy scale $\Lambda_M > 1\,{\rm TeV}$, and regulated photon and electron propagators. The problem of guaranteeing the gauge invariance of QED was solved in previous work~\cite{MoffatWoodard1991}. The gauge invariance of Yang-Mills theory with non-Abelian gauge invariance was extended by Kleppe and Woodard~\cite{WoodardKleppe1992}. In the following, we will concentrate on the electroweak (EW) and Higgs boson sectors.

\section{Finite Quantum Field Theory}

When we use the Minkowski spacetime, we adopt the Minkowski metric convention, $\eta_{\mu\nu}={\rm diag}(+1,-1,-1,-1)$, and unless otherwise stated we set $\hbar=c=1$. Let us consider the example of four-dimensional real scalar field theory with the Euclidean signature action:
\be
S_\phi=\int d^4x\biggl(-\frac{1}{2}\phi\Box\phi+\frac{1}{2}m^2\phi^2+\frac{1}{4!}\lambda\phi^4\biggr),
\ee
where $\Box=\partial_\mu\partial_\mu$. We have chosen the self-interaction $\lambda\phi^4$ but we could have chosen any polynomial power of $\phi$. The bare Euclidean propagator in standard local QFT is given by
\be
\Delta(x)=\frac{1}{4\pi^2}\biggl(\frac{m}{x}\biggr)K_1(mx),
\ee
where $x^2=x_\mu x_\mu$, $x=\sqrt{x_\mu x_\mu}$ and $K_1(z)$ is the modified Bessel function of the second kind. The short-distance expansion is
\be
\Delta(x)=\frac{1}{4\pi^2}\biggl[\frac{1}{x^2}+\frac{1}{4}m^2\ln(m^2x^2)+\frac{m^2}{4}(1-2\psi(2))+r(mx)\biggr],
\ee
where $\psi$ is the digamma function, $\psi(2)=1-\gamma$, $\gamma$ is Euler's constant and $r(mx)\sim mx$ as $mx\rightarrow 0$.

The regularized position propagator $\tilde\Delta(x-x')$ in Minkowski spacetime is the Green's function $G(x,x')$ for the Klein-Gordon equation:
\be
\label{Regpropagator}
(\Box_x+m^2)\tilde\Delta(x-x')\equiv {\cal E}(x-x')=-\frac{1}{4\pi^2\Lambda_x^4}\exp\biggl(-(x-x')^2/2\Lambda_x^2\biggr).
\ee
In the limit $\Lambda_x\rightarrow 0$, we obtain the local Klein-Gordon equation:
\be
(\Box_x+m^2)\Delta(x-x')=-\delta^4(x-x').
\ee
A Fourier transform of (\ref{Regpropagator}) leads to
\be
(-p^2+m^2)\tilde\Delta(P)\equiv{\cal E}(P)=\int_{-\infty}^\infty d^4y
\exp\biggl(-iP\cdot y/\hbar\biggr)\exp\biggl(-y^2/2\Lambda_y^2\biggr),
\ee
where $y=x-x'$ and $P=p-p'$. We obtain for ${\cal E}(P)$:
\be
{\cal E}(P)=\int_{-\infty}^{\infty}d^4y\exp\biggl[-\frac{1}{2}\biggl(y/\Lambda_y+iP\Lambda_y/\hbar\biggr)^2\biggr]
\exp\biggl(-P^2\Lambda_y^2/2\hbar^2\biggr).
\ee
We now define $\Lambda_P$ by the relation:
\be
\Lambda_y\Lambda_P=\hbar,
\ee
which gives to within a normalization factor:
\be
{\cal E}(P)=\exp\biggl(-P^2/2\Lambda_P^2\biggr).
\ee
The probability distribution is given by
\be
{\cal P}(p)=|{\cal E}(p)|^2.
\ee
We display the probability distribution in Fig. 1. This shows that the distribution function has a value for the momentum $P$ which differs from zero by at most $\Lambda_P$. The uncertainty $\Lambda_y$ in position space $y$, and the uncertainty $\Lambda_P$ in momentum space can be related to the Heisenberg uncertainty principle:
\be
\Lambda_y\Lambda_P\geq\hbar.
\ee
In the limit $\Lambda_P\rightarrow 0$ we have
\be
\int_{-\infty}^{\infty}d^4P{\cal E}(P)\rightarrow\int_{-\infty}^{\infty}d^4P\delta^4(P)=1.
\ee
Thus, we see that the strictly local QFT determined by $\delta^4(x-x')$ in position space and $\delta^4(p-p')$ in momentum space {\it corresponds to zero uncertainty} in the position and momentum spaces.

In the collision of two particles, we cannot according to the quantum uncertainty know simultaneously the precise values of the $x$ or $p$ values for the vertex or propagator distributions. In the strictly local QFT, the propagator and vertex distribution is chosen so that $\Lambda_P\rightarrow\infty$ leading to
$\exp\biggl(-\frac{p^2}{2\Lambda_p^2}\biggr)=1$. The distributions ${\cal E}(y)$ and ${\cal E}(P)$ are entire functions in Euclidean position and momentum spaces~\cite{Moffat1990,MoffatWoodard1991,Moffat2011}. We have for a free particle, $p_\mu\rightarrow -i\frac{\partial}{\partial x_\mu}$ and $\exp(\Box/2\Lambda_x^2)\rightarrow\exp(-p^2/2\Lambda_P^2)$.

\begin{figure}
\centering\includegraphics[scale=1]{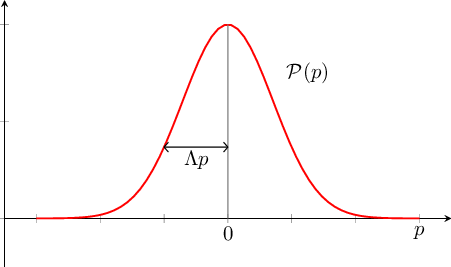}\\
\caption{The probability distribution corresponding to the Gaussian, entire function distribution ${\cal E}(p)$. The value of $p$ is almost certain to be found within the $p$ range $\Lambda_p$. In the limit $\Lambda_p\rightarrow 0$ the peak at the origin becomes infinitely high and narrow and the distribution ${\cal E}(p)$ becomes a delta function $\delta(p)$.}{\label{fig.P-plot}}
\end{figure}

Let us consider the question of causality in our finite QFT. We define the nonlocal field operator:
\be
\tilde\phi(x)=\int d^4x'{\cal F}(x-x')\phi(x')={\cal F}({\cal E}(x))\phi(x),
\ee
where $\phi(x)$ is the local field operator and ${\cal F}(x)={\cal F}({\cal E})$ is an entire function distribution operator. We have
\be
\phi(x)=\int\frac{d^3p}{(2\pi)^3}\frac{1}{\sqrt{2E_p}}[a_p\exp(-ip\cdot x)+a_p^\dagger\exp(ip\cdot x)],
\ee
where $a_p$ and $a_p^\dagger$ are creation and annihilation operators. In the Heisenberg picture, the regularized amplitude for a particle to propagate from $x$ to $x'$ is
\be
\langle 0|\tilde\phi(x)\tilde\phi(x')|0\rangle=\tilde\Delta(x-x')=\int\frac{d^3p}{(2\pi)^3}\frac{1}{\sqrt{2E_p}}
\exp(-ip\cdot(x-x')){\cal F}(x-x').
\ee

We now consider the c-number commutator $[\tilde\phi(x),\tilde\phi(x')]=\langle 0|[\tilde\phi(x),\tilde\phi(x')]|0\rangle$:
\be
[\tilde\phi(x),\tilde\phi(x')]=\int\frac{d^3p}{(2\pi)^3}\frac{1}{\sqrt{2E_p}}[\exp(-ip\cdot (x-x'))-\exp(ip\cdot(x-x'))]
{\cal F}(x-x')=\tilde\Delta(x-x')-\tilde\Delta(x'-x).
\ee
Each term is separately Lorentz invariant.  When $(x-x')^2 < 0$, we can perform a Lorentz transformation on the second term yielding $(x-x')\rightarrow -(x-x')$. The two terms become equal and cancel to give zero, whereby causality is preserved:
\be
[\tilde\phi(x),\tilde\phi(x')]=0,\quad (x-x')^2 < 0.
\ee
Thus, we conclude that no measurement in the nonlocal finite QFT can affect another measurement outside the light cone. It is accepted that the phenomenon of quantum entanglement is a generic nonlocal behavior of quantum mechanics but that there is no violation of causality. No information is exchanged at the speed of light, so that causality is not violated. {\it We have demonstrated that micro-causality is preserved in nonlocal finite QFT.}

The action is written as a free part plus an interaction part:
\be
S_\phi=S_F(\phi)+S_I(\phi),
\ee where
\be
S_F(\phi)=\int d^4x\biggl(\frac{1}{2}\phi(x){\cal K}\phi(x)\biggr).
\ee
We define from the kinetic operator ${\cal K}$ the distribution operator ${\cal E}$:
\be
\label{Edistr}
{\cal E}=\exp\bigg(\frac{{\cal K}}{2\Lambda^2}\biggr).
\ee

The Feynman rules for the finite QFT follow as extensions of the local standard QFT. For every internal line in a diagram can be connected to a regulated propagator:
\be
\label{propagatorReg}
i\tilde\Delta=\frac{i{\cal E}^2}{{\cal K}}=i\int\frac{d\tau}{\Lambda^2}\exp\biggl(\tau\frac{\cal K}{\Lambda^2}\biggr).
\ee
We have used the Schwinger proper time method to determine the propagator.

In the formulation of finite QED~\cite{MoffatWoodard1991}, an additional auxiliary propagator was introduced:
\be
\label{auxpropagator}
-i\hat\Delta=\frac{i(1-{\cal E}^2)}{{\cal K}}=-i\int_0^1\frac{d\tau}{\Lambda^2}\exp\biggl(\tau\frac{{\cal K}}{\Lambda^2}\biggr).
\ee
The auxiliary propagator $\hat\Delta$ does not possess poles and does not have particles. Tree order amplitudes such as Compton scattering amplitudes are identical to their local QFT counterparts. The tree amplitudes such as Compton amplitudes are the sum of (\ref{propagatorReg}) and (\ref{auxpropagator}), and this sum gives the standard local propagator and tree graphs and they are free of unphysical couplings.

In the following, we will prescribe a more general formulation involving two energy scales $\Lambda_M$ and $\Lambda_H$. The $\Lambda_M$ controls the energy scale at which the entire function distribution begins to take effect in scattering processes and quantum loops involving the bosons, quarks and leptons of the SM, while $\Lambda_H$ controls the energy scale of the Higgs boson quantum loops.

The bare regularized Feynman propagator in Euclidean momentum space is
\be
\label{Feynmanpropagator}
i\tilde\Delta_F(p)=i\frac{\exp(-(p^2+m^2)/\Lambda_p^2)}{p^2+m^2}.
\ee
The fermion Feynman propagator is given by
\be
i\tilde S_F(p)=i\frac{({\slashed p}+m)\exp(-(p^2+m^2)/\Lambda_p^2)}{p^2+m^2},
\ee
where ${\slashed p}=\gamma^\mu p_\mu$. The S-matrix for scattering amplitudes satisfies the Cutkosky rules~\cite{Cutkosky} and unitarity to all orders of perturbation theory~\cite{MoffatWoodard1991,Tomboulis}.

To quantize the finite QFT, we use the functional formalism based on the Feynman path integral~\cite{MoffatWoodard1991}. For gauge invariant QED and QCD the problem of quantization amounts to finding an acceptable measure factor, which makes the functional formalism invariant under the classical gauge interactions. This means that the measure factor in the path integral obeys the same restrictions as the classical one, namely, manifest Poincar\'e invariance, exponential suppression in Euclidean momentum space, reality for real momenta, and analyticity in momentum variables throughout the complex plane.

The perturbation theory can be performed in Minkowski space, while the loop integrations are done by Wick rotation or by an analytic continuation to Euclidean momentum space. The whole formalism can be extended to QED, QCD, weak interactions and quantum gravity. The quantum loops can be described either by internal regulated propagators or by distribution coupling constant vertices ${\tilde g}_i(p)=g_i{\cal E}(p)$.

The issue of the conservation of the Noether axial current:
\be
J^\mu_5=\bar\psi\gamma^\mu\psi
\ee
in the finite QED have been considered in the past~\cite{MoffatWoodard1991,ClaytonDemopolousMoffat}. It was demonstrated that the QED anomalous results for triangle graphs and the decay rate $\pi^0\rightarrow 2\gamma$ are correctly produced. The finite QFT can resolve the problem of fermion doubling in QCD calculations, a problem that is produced by demanding standard infinite renormalization of chiral fermion amplitudes. Another consequence of the finite QFT is the removal of the Landau pole in QED and scalar field theory~\cite{Moffat1990,Ghoshal}.

Superstrings have demonstrated that they can satisfy both ultraviolet finiteness and unitarity, and that superstrings can therefore solve the lack of renormalizability of quantum gravity. The finite QFT proposal we are promoting makes the QFT ultraviolet finite and unitarity in four-dimensional theory, without invoking string theory's one-dimensional strings and 10 and 11-dimensional spacetimes. The compactification of string theory higher dimensions leads to the very large landscapes of particle theories. The ultraviolet finiteness of quantum loop amplitudes in string theory is a consequence of the fact that vertices in Euclidean momentum space are suppressed by factors $\exp(-\alpha' p^2)$, where $\alpha'$ is the superstring tension. These factors are infinite derivative entire functions of momentum, which we utilize to make our finite QFT for point particles in four dimensions.

For vector and tensor theories such as QED, QCD and QG, the entire function distribution can break gauge invariance and unitarity. However, it was proved that gauge invariance and unitarity can be restored by higher order interactions~\cite{MoffatWoodard1991}. This restoration of gauge invariance and unitarity and the unitarity Cutkosky rules~\cite{Cutkosky} was made possible by invoking a nonlinear gauge invariance,  which agrees with the usual local symmetry on shell but is larger off shell.

Let us consider the Callan-Symanzik equations~\cite{Callan,Symanzik}. In finite QFT theory, the equations for the regularized amplitudes $\Gamma^{(n)}(x-x')$ are
\be
\biggl[\mu \frac{\partial}{\partial\mu}+\beta({\tilde g}_i)\frac{\partial}{\partial {\tilde g}_i}
-2\gamma({\tilde g})\biggr]\Gamma^{(n)}=0,
\ee
where $\mu$ is the renormalization group mass scale and ${\tilde g}_i$ are the running coupling constants associated with diagram vertices. The correlation functions will satisfy this equation for the n-th order $\Gamma^{(n)}$ for the Gell-Mann-Low functions $\beta({\tilde g}_i)$ and the anomalous dimensions in nth-loop order.

\section{The Electroweak Lagrangian}

The theory introduced here is based on the $SU(3)_c\times SU_L(2)\times U_Y(1)$ Lagrangian that includes leptons and quarks with the color degree of freedom of the strong interaction group $SU_C(3)$.  Let us now consider the EW sector. {\it We assume that the $SU(2)_L\times U(1)_Y$ local gauge group symmetry is broken by the massive boson and fermion particles}. The EW model Lagrangian is given by
\begin{align}
\label{Lagrangian}
{\cal L}_{\rm EW}=\sum_{\psi_L}\tilde\psi_L\biggl[\gamma^\mu\biggl(i\partial_\mu - \frac{1}{2}{\tilde g}\tau^aW^a_\mu - {\tilde g}'\frac{Y}{2}B_\mu\biggr)\biggr]\psi_L\nonumber\\
+\sum_{\psi_R}\tilde\psi_R\biggl[\gamma^\mu\biggl(i\partial_\mu - {\tilde g}'\frac{Y}{2}B_\mu\biggr)\biggr]\psi_R -\frac{1}{4}B^{\mu\nu}B_{\mu\nu}\nonumber\\
-\frac{1}{4}W_{\mu\nu}^aW^{a\mu\nu} + {\cal L}_M + {\cal L}_{m_f}.
\end{align}
The ${\tau}'s$ are the usual Pauli spin matrices and $\psi_L$ denotes a left-handed fermion (lepton or quark) doublet, and the $\psi_R$ denotes a right-handed fermion singlet. The fermion fields (leptons and quarks) have been written as $SU_L(2)$ doublets and U(1)$_Y$ singlets, and we have suppressed the fermion generation indices. We have $\psi_{L,R}=P_{L,R}\psi$, where $P_{L,R}=\frac{1}{2}(1\mp\gamma_5)$. Moreover, ${\tilde g}(p^2)=g{\cal E}(p^2/\Lambda_M^2)$ and
${\tilde g}'(p^2)=g'{\cal E}(p^2/\Lambda_M^2)$.

We also have the Lagrangian for the neutral scalar Higgs boson:
\be
{\cal L}_{\rm Higgs}=\biggl|\biggr(i\partial_\mu -\frac{1}{2}{\tilde g_H}\tau^a W_\mu^a-{\tilde g_H}'\frac{Y}{2}B_\mu\biggr)\phi\biggr|^2
+\frac{1}{2}m_H^2\phi^2,
\ee
where $\phi$ is the isoscalar neutral Higgs field, ${\tilde g_H}(p^2)=g_H{\cal E}(p^2/\Lambda_H^2)$ and ${\tilde g'_H}(p^2)=g'_H{\cal E}(p^2/\Lambda_H^2)$.  The photon-fermion Lagrangian is
\be
L_{\rm QED}=\sum_{\psi_L}\tilde\psi_L\biggl[\gamma^\mu\biggl(i\partial_\mu - \frac{1}{2}{\tilde e}\biggr)A_\mu\biggr]\psi_L
+\sum_{\psi_R}\tilde\psi_L\biggl[\gamma^\mu\biggl(i\partial_\mu - \frac{1}{2}{\tilde e}\biggr)A_\mu\biggr]\psi_R -\frac{1}{4}F^{\mu\nu}F_{\mu\nu}+{\cal L}_{m_f},
\ee
where ${\tilde e}(p^2)=e{\cal E}(p^2/\Lambda_M^2)$. Moreover,
\begin{equation}
\label{Bequation}
B_{\mu\nu}=\partial_\mu B_\nu-\partial_\nu B_\mu,
\end{equation}
\begin{equation}
W^a_{\mu\nu}=\partial_\mu W_\nu^a-\partial_\nu W_\mu^a-{\tilde g}f^{abc}W_\mu^bW_\nu^c.
\end{equation}
and
\be
F_{\mu\nu}=\partial_\mu A_\nu-\partial_\nu A_\mu.
\ee
The quark and lepton fields, the boson fields $W^a_\mu$ and $B_\mu$ and the Higgs and photon fields satisfy microcausality.

The $\Lambda_M$ and $\Lambda_H$ are energy scales, which are measurable parameters in the model and ${\cal E}$ is an {\it entire function} of $\Box=\partial^\mu\partial_\mu$ or $p^2$. The coupling functions ${\tilde g}$, ${\tilde g}'$, ${\tilde g}_H$ and ${\tilde e}$ are scalars under a Lorentz transformation.

The Lagrangian for the vector boson mass terms is
\begin{equation}
{\cal L}_M=\frac{1}{2}M^2_WW^{a\mu} W^a_\mu + \frac{1}{2}M^2_BB^\mu B_\mu,
\end{equation}
and the fermion mass Lagrangian is
\begin{equation}
\label{fermionmass}
{\cal L}_{m_f}=-\sum_{\psi_L^i,\psi_R^j}m_{ij}^f(\tilde\psi_L^i\psi_R^j + \tilde\psi_R^i\psi_L^j),
\end{equation}
where $M_W$, $M_B$ and $m_{ij}^f$ denote the boson and fermion masses, respectively. Eq.(\ref{fermionmass}) can incorporate massive neutrinos and their flavor oscillations. The mass Lagrangians explicitly break $SU(2)_L\times U(1)_Y$ gauge symmetry.

The $SU(2)$ generators satisfy the commutation relations
\begin{equation}
[T^a,T^b]=if^{abc}T^c,~~~~~\mathrm{with}~~~~~T^a=\frac{1}{2}\sigma^a.
\end{equation}
Here, $\sigma^a$ are the Pauli spin matrices and $f^{abc}=\epsilon^{abc}$. The fermion--gauge boson interaction terms are contained in
\begin{equation}
L_I=-i{\tilde g}J^{a\mu}W_\mu^a-i{\tilde g}'J_Y^\mu B_\mu,
\end{equation}
where the $SU(2)$ and hypercharge currents are given by
\begin{equation}
J^{a\mu}=-i\sum_{\psi_L}\tilde{\psi}_L\gamma^\mu T^a\psi_L,~~~~~\mathrm{and}~~~~~J_Y^\mu=-i\sum_\psi\frac{1}{2}Y\tilde\psi\gamma^\mu\psi,
\end{equation}
respectively. The last sum is over all left and right-handed fermion states with hypercharge factors $Y=2(Q-T^3)$ where $Q$ is the electric charge.

We diagonalize the charged sector and perform mixing in the neutral boson sector. We write
$W^\pm=\frac{1}{\sqrt{2}}(W^1\mp iW^2)$ as the physical charged vector boson fields.
In the neutral sector, we can mix the $Z_\mu$ field in the usual way:
\begin{equation}
Z_\mu=\cos\theta_wW_\mu^3-\sin\theta_wB_\mu.
\label{eq:2.35}
\end{equation}
We define the usual relations
\label{sintheta}
\begin{equation}
\sin^2\theta_w=\frac{g'^2}{g^2+g'^2}~~~\mathrm{and}~~~\cos^2\theta_w=\frac{g^2}{g^2+g'^2}.
\end{equation}

If we identify the resulting $A_\mu$ field with the photon, then we have the unification condition:
\begin{equation}
e=g\sin\theta_w=g'\cos\theta_w
\end{equation}
and the electromagnetic current is
\begin{equation}
J_\mathrm{em}^\mu=J^{3\mu}+J_Y^\mu.
\end{equation}
The neutral current is given by
\begin{equation}
J_\mathrm{NC}^\mu=J^{3\mu}-\sin\theta_wJ_\mathrm{em}^\mu,
\end{equation}
and the fermion-boson interaction terms are given by
\begin{equation}
L_I=-\frac{\tilde g}{\sqrt{2}}(J_\mu^+W^{+\mu}+J_\mu^-W^{-\mu})-{\tilde g}\sin\theta_wJ_\mathrm{em}^\mu A_\mu-\frac{\tilde g}{\cos\theta_w}
J_\mathrm{NC}^\mu Z_\mu.
\end{equation}

Gauge invariance is important for the QED sector, $U_{\rm em }(1)$, for it leads to a consistent quantization of QED calculations by guaranteeing that the Ward-Takahashi identities are valid. The gauge invariant QED is constructed so that classical tree graphs are strictly local~\cite{MoffatWoodard1991}. Quantization of the Proca massive vector boson EW sector of $SU(2)\times U(1)$ is physically consistent even though the $SU(2)\times U(1)$ gauge symmetry is dynamically broken by masses.

\section{Broken $SU(2)\times U(1)$ symmetry}

The SM formulation of the weak interaction sector is initiated with an $SU(2)\times U(1)$ gauge invariant symmetry with all boson and fermion masses equal to zero. This allows for an application of standard perturbative renormalization theory. To resurrect the particle masses, the Higgs field invokes a spontaneous breaking of the vacuum symmetry, while retaining the Lagrangian density $SU(2)\times U(1)$ symmetry. The spontaneous breaking of the $SU(2)\times U(1)$ symmetry is invoked by the Higgs field doublet:
\be
\phi_H=\frac{1}{\sqrt{2}}\begin{pmatrix}0\\v+h(x)\end{pmatrix},
\ee
where $v=246$ GeV and $h(x)$ is the neutral Higgs particle field. Upon substitution of $\phi_H$, the Yukawa Lagrangian becomes:
\be
{\cal L}_Y=\frac{g_f}{\sqrt{2}}v(\bar\psi_L\psi_R+\bar\psi_R\psi_L)+\frac{g_f}{\sqrt{2}}(\bar\psi_L\psi_R+\bar\psi_R\psi_L)h.
\ee
We choose the coupling fermion constants $g_f$ so that
\be
m_f=\frac{1}{\sqrt{2}}{g_fv},
\ee
generating the fermion masses:
\be
{\cal L}_Y=m_f\bar\psi\psi+\frac{m_f}{v}\bar\psi\psi h.
\ee
This predicts that the fermion coupling constants are proportional to the fermion masses. For the vector bosons $W$ and $Z$ the spontaneous symmetry breaking produces the results:
\be
M_W=\frac{1}{2}vg,\quad M_Z=\frac{1}{2}v\sqrt{g^2+g^{'2}},
\ee
and for the bosons the coupling constants are proportional to the $W$ and $Z$ masses. The experimental evidence for these predictions are good for the heavy $W$ and $Z$ bosons. However, for the third generations of quarks and leptons they are at best 20-30\%.

In our broken $SU(2)\times U(1)$ model, we can model the intrinsic broken symmetry by the breaking scale
$b\equiv E_{\rm EW}=\biggl(\frac{1}{\sqrt2}{G_F}\biggr)^{1/2}=246\,{\rm GeV}$, while maintaining only the true vacuum with $v=\langle 0|\phi_H|0\rangle=0$~\cite{Moffat2011}.

The ultimate test of the Higgs field spontaneous symmetry breaking of the vacuum model and the Higgs mechanism in the SM is the experimental determination of the shape of the potential $V(\phi)$ given by (\ref{GinsbergLandaupotential}). This can be done by a measurement of the triple or quartic coupling of the neutral Higgs boson field. This will be a difficult measurement to perform. It can be done by measuring the far off-shell decay of the Higgs boson into two Higgs bosons by a future circular $e^+$ and $e^-$ collider with an energy greater than about 500 GeV.

\section{Massive Boson Propagators and Perturbative Unitarity}

We construct a QFT that is UV complete in perturbation theory and does not violate unitarity of scattering amplitudes. We do not attempt to generate masses of the fermions and bosons as was done by the spontaneous symmetry breaking of the vacuum in the standard Higgs model, or as was done in the non-local regularized EW model~\cite{Moffat1991}.

The lack of renormalizability and the violation of unitarity of the weak interactions with an intermediate, massive vector boson is caused by the {\it longitudinal} component of the polarization vector, $\epsilon^\mu(p,\lambda=0)$~\cite{Moffat2011}. In a frame in which $p^\mu=(p^0,0,0,|{\bf p}|)$, the transverse polarization vectors $\epsilon^\mu(p,\lambda=\pm 1)$ involve no momentum dependence; it is carried solely in the longitudinal polarization vector. We write the longitudinal polarization vector for the $W$ meson as
\begin{equation}
\epsilon(p,\lambda=0)=\frac{p^\mu}{M_W}+\frac{M_W}{(p^0+|{\bf p}|)}(-1,{\hat{\bold p}}).
\end{equation}
This tends at high energy to $p^\mu/M_W$.

To solve the problem of avoiding infinite renormalization of weak interactions, we invoke energy dependent couplings at Feynman diagram vertices connecting a longitudinal $W_L$ to fermions in loop diagrams:
\begin{equation}
\label{equation}
{\tilde g}_L(p^2)=g{\cal E}_L(p^2/\Lambda_W^2),\quad {\tilde g}'_L(p^2)={\tilde g}'{\cal E}_L(p^2/\Lambda_W^2),\quad {\tilde g}_T(p^2)=g,
\end{equation}
where ${\cal E}_L$ pertains to the part of the vertex connecting a longitudinal $W_L$ to fermions. The entire function ${\cal E}_L$ vanishes rapidly enough at high energies to cancel any unitarity violation in scattering amplitudes for longitudinally polarized $W$s for $p^2 > \Lambda_M^2$. Alternatively, we can keep the local vertices but make the propagators within internal loop diagrams regulated by the entire function distribution ${\cal E}$.

The ${\cal E}_L(p^2/\Lambda_M^2)$ is an {\it entire} function for complex $p^2$ which satisfies on-shell ${\cal E}_L(0)=1$. This allows us to obtain a Poincar\'e invariant, finite and unitary perturbation theory. Such entire functions are analytic (holomorphic) in the finite complex $p^2$ plane. They must possess a pole or an essential singularity at the point at infinity, for otherwise by Liouville's theorem they are constant. Because they contain no poles for finite $p^2$, {\it they do not produce any unphysical particle poles and unwanted degrees of freedom.} corresponding to poles on the physical Riemann sheet. Provided that the vertex couplings ${\tilde g}_L(p^2)$ and ${\tilde g}'_L(p^2)$ decrease fast enough for $p^2 > \Lambda_M^2$ in Euclidean momentum space, the problem is removed of the lack of renormalizability of our minimal EW action containing only the observed twelve quarks and leptons, the $W$ and $Z$ bosons, the Higgs boson and the massless photon.

The $W^\pm$ and $Z^0$ mass Lagrangian is given by
\begin{equation}
{\cal L}_M=M_W^2W^+_\mu W^{-\mu} + \frac{1}{2}M_Z^2Z_\mu Z^\mu.
\end{equation}

There is no known fundamental motivation for choosing the SM model Landau-Ginsburg-Higgs potential (\ref{GinsbergLandaupotential}) with $\mu^2 < 0$. We could add an additional contribution $\lambda'(\phi^\dagger\phi)^3$ to the potential (\ref{GinsbergLandaupotential}) or even higher order polynomials in $\phi$. The fact that such higher-dimensional operators render the EW model non-renormalizable would not justify their lack of inclusion in our UV finite model. The quark and lepton masses, the $W$ and $Z$ masses and the Higgs boson mass are the physical masses in the propagators. We circumvent the problem of the lack of renormalizability of our model by damping out divergences with the coupling vertices ${\tilde g}(p^2)$, ${\tilde g}'(p^2)$, ${\tilde g}_H$ and ${\tilde e}(p^2)$. We emphasize that our energy scale parameters $\Lambda_M$ and $\Lambda_H$ are not naive cutoffs. The entire function property of the coupling vertices guarantees that the model suffers no violation of unitarity, Poincar\'e invariance or the gauge invariance of the QCD and QED sectors.

Given the experimentally determined $W$ and $Z$ masses and the weak angle $\cos\theta_w$, we obtain the relation satisfied at the effective tree graph level:
\be
\rho=\frac{M_W^2}{M_Z^2\cos^2\theta_w}=1.
\ee

\section{Quantization of Proca Massive Bosons}

Let us consider the canonical quantization of the massive Proca fields $W^{a\mu}$ and $B^\mu$~\cite{Moffat2011}. The Proca fields have only three independent dynamical degrees of freedom. This can be seen from the equation of motion for the $B_\mu$ field:
\begin{equation}
\partial_\mu B^{\mu\nu}+M^2_BB^\nu=J_Y^\nu,
\end{equation}
which can be written as
\begin{equation}
\Box B^\nu-\partial^\nu(\partial_\mu B^\mu)+M_B^2 B^\nu=J_Y^\nu.
\end{equation}
The four-divergence of this equation gives
\begin{equation}
\partial_\nu B^\nu=\frac{1}{M_B^2}\partial_\nu J^\nu_Y.
\end{equation}
The source current $J^\nu_Y$ need not be conserved for the Proca field. However, if we assume that it is, $\partial_\nu J^\nu_Y=0$, then we have that
\begin{equation}
\label{Lorenz}
\partial_\nu B^\nu=0,
\end{equation}
is automatically satisfied. The Lorenz condition becomes a constraint equation for the Proca field, making the $B^0$ a dependent variable. We have the Proca equation for $J_Y^\nu=0$:
\begin{equation}
\partial_\mu B^{\mu0}+M_B^2B^0=0.
\end{equation}
This yields the equation
\begin{equation}
\label{Bconstraint}
B^0=-\frac{1}{M_B^2}\partial_iB^{i0},
\end{equation}
which shows that $B^0$ is a dependent quantity and not an independent dynamical degree of freedom.
The Hamiltonian for the $B^\mu$ field is given by
\begin{equation}
H_B=\int d^3x\frac{1}{2}\biggl[(B^{i0})^2+(B^{ij})^2+M_B^2(B^{i0})^2+\frac{1}{M_B^2}(\partial_iB^{i0})^2\biggr].
\end{equation}

Let us now turn to the non-Abelian gauge field $W^a_\mu$. The covariant derivative operator is given by
\begin{equation}
D^\mu W^a_{\mu\nu}\equiv \partial^\mu W^a_{\mu\nu}+igf^{abc}W^{b\mu}W^c_{\mu\nu}.
\end{equation}
The equations of motion are
\begin{equation}
D_\mu W^{a\mu\nu}+M_W^2W^{a\nu}=J^{a\nu}.
\end{equation}
Taking $J^{a\nu}=0$ and $\nu=0$ gives
\begin{equation}
\label{Wconstraint}
W^{a0}=-\frac{1}{M_W^2}D_iW^{ai0}.
\end{equation}
As with the $U(1)$ Abelian field $B^\mu$ the $W^{a0}$ is not an independent dynamical degree of freedom. The Hamiltonian for the $W^{a\mu}$ field is
\begin{equation}
H_W=\int d^3x\frac{1}{2}\biggl[(W^{ai0})^2+(W^{aij})^2+M_W^2(W^{ai0})^2+\frac{1}{M_W^2}(D_iW^{ai0})^2\biggr].
\end{equation}
We observe that the canonical quantization of the local boson field operators $W^{a\mu}$ and $B^\mu$ leads to positive energy ghost-free Hamiltonians. For the nonlocal field operators ${\tilde W}^{a\mu}$ and ${\tilde B}^\mu$ a similar Proca ghost-free and consistent quantization can be formulated.

A covariant quantization of the Proca fields can be derived by imposing the second class constraints on the field operators using (\ref{Bconstraint}) and (\ref{Wconstraint}) as operator constraints~\cite{Dirac}. We have for the local $W^{a\mu}$ and $B^\mu$ fields the equal time commutation relations:
\begin{align}
[B^{i0}({\bf x},t),B^0({\bf x}',t))&=\frac{i}{M_B^2}{\bf\nabla}^i\delta^3({\bf x}-{\bf x}'),\nonumber\\
[B^0({\bf x},t),B^0({\bf x}',t)]&=0,
\end{align}
and
\begin{align}
[W^{ai0}({\bf x},t),W^{b0}({\bf x}',t)]&=\delta^{ab}\frac{i}{M_W^2}{\bf\nabla}^i\delta^3({\bf x}-{\bf x}'),\nonumber\\
[W^{a0}({\bf x},t),W^{b0}({\bf x}',t)]&=0.
\end{align}
These commutation relations can be extended to the nonlocal field operators ${\tilde B}^\mu$ and ${\tilde W}^{a\mu}$.

\section{Perturbative Unitarity and the Higgs Boson}

The standard EW model violates unitarity in scattering processes that involve longitudinally polarized vector bosons without the Higgs particle. The scattering of two longitudinally polarized vector bosons $W_L$ results in a divergent term proportional to the center-of-mass energy squared $s$. A less rapid divergence, proportional to $\sqrt{s}$, occurs when fermions annihilate into a pair of $W_L$ vector bosons. The tree-level processes involving the Higgs boson cancel these divergences.

The scattering amplitude matrix elements for the process $W^+_L + W^-_L\rightarrow W^+_L + W^-_L$ is given by~\cite{Aitchison}:
\begin{equation}
i{\cal M}_{W_L}=ig^2\left[\frac{\cos\theta+1}{8M_W^2}s+{\cal O}(1)\right],
\label{eq:MM2}
\end{equation}
where $\theta$ is the scattering angle. This result clearly violates perturbative unitarity for large $s$ for $\Lambda_M < 1-14$ TeV. This behavior is corrected by the addition of the $s$-channel Higgs exchange in the high-energy limit:
\begin{equation}
i{\cal M}_H=-ig^2\left[\frac{\cos\theta+1}{8M_W^2}s+{\cal O}(1)\right].
\end{equation}
This cancels out the bad behavior in (\ref{eq:MM2}). Our finite QFT model {\it still requires a Higgs boson to guarantee perturbative unitarity for} $\Lambda_M < 1-14$ TeV.

We have postulated that all Feynman tree graphs in our Poncar\'e and unitary model possess a vertex operator
${\cal V}(p^2)=g_i{\cal E}(p^2,\Lambda_M)$ for $\Lambda_M >14\, {\rm TeV}$. This guarantees that all tree graphs for low energies satisfy locality and microcausality with delta-function vertex and propagator distributions, corresponding to point-like interactions. The Feynman loop graphs will have the vertex operator
${\cal V}(p^2)=g_i{\cal E}(p^2,\Lambda_{M,H})$ or regulated propagators, guaranteeing that the loop graphs are finite to all orders of perturbation theory,  This removes the need for infinite renormalization, although the perturbative theory will have {\it finite} renormalization of mass and charge. The Higgs-fermion loops will be controlled by the energy scale $\Lambda_H\lesssim 1\,{\rm TeV}$. The tree graph scattering amplitudes and tree graph decay amplitudes obey the same rules as in the SM without incurring infinite divergences.

We will identify an EW energy scale:
\be
\label{EWweakscale}
E_{\rm EW}=\biggl(\frac{1}{\sqrt2}{G_F}\biggr)^{1/2}=246\,{\rm GeV},
\ee
where $G_F=1.16638\times 10^{-5}\,{\rm GeV}^{-2}$ is Fermi's coupling constant. In the SM $v=v_{\rm EW}$ is the vacuum expectation value of the Higgs field and the Higgs mechanism value $M_W=\frac{1}{2}vg$ is used together with $\sqrt{2}g^2/8M_W^2=G_F$ to obtain (\ref{EWweakscale}). In our finite QFT EW model, the potential $V(\phi)$ has only one minimum  $\langle 0|\phi_H\|0 \rangle =0$ corresponding to the true vacuum. With our tree graph rules, we obtain the standard decay rate for the decay of the Higgs boson into fermions:
\be
\Gamma(H\rightarrow f{\tilde f})=\frac{N_C}{8\pi E_{\rm EW}^2}m_f^2m_H\beta_f^3,
\ee
where $\beta_f=\sqrt{1-4m_f^2/m_H^2}$ and $N_C$ denotes the number of quarks. The tree graph decay predictions of the Higgs into lighter bosons and fermions will agree with CERN experiments.

\section{Radiative Corrections}

Radiative corrections alter the $\rho$ parameter determining the relative strength of the neutral to charged currents $J^\mu_ZJ_{\mu Z}/J^{\mu +}J_\mu^{-}$:
\begin{equation}
\rho=\rho_{(0)} + \Delta\rho_{(1)},
\end{equation}
where $\rho_{(0)}=1$ and $\Delta\rho_{(1)}$ denotes the one-loop correction dominated by the top quark. For low-energies: $p^2 < \Lambda^2_M$ with $\Lambda_M > 1$ TeV we get
\begin{equation}
\label{deltarho}
\Delta\rho_{(1)}\sim\frac{3G_FM_W^2}{8\pi^2\sqrt{2}}\biggl(\frac{m_t^2}{M_W^2} + \frac{5}{6}\biggr).
\end{equation}
There is an extra contribution to (\ref{deltarho}) coming from the Higgs particle:
\begin{equation}
\label{Higgsdelta}
\Delta\rho_{H(1)}\sim -\frac{3G_FM_W^2}{8\pi^2\sqrt{2}}\biggl[\biggl(\frac{M_Z^2}{M_W^2}-1\biggr)\ln\biggl(\frac{m_H^2}{M_W^2}\biggr)\biggr],
\end{equation}
where $m_H=125$ GeV, $M_W=80.39$ GeV and  $M_Z=91.18$ GeV. We have $M_Z^2/M_W^2-1=\sin^2\theta_w/\cos^2\theta_w$ and we obtain in our model from (\ref{deltarho}) the value $\rho\sim 1.01$.

For the Higgs mass, $m_H=125\,{\rm GeV}$, the non-oblique radiative Higgs corrections are not important. An example of this is $Z\rightarrow b +{\tilde b}$ decay. The Higgs loop corrections for this process for the decay of the Higgs are proportional to the coupling $\lambda_b\sim \sqrt{2}m_b$ and are negligible~\cite{Logan}. Therefore, there is no need for these non-oblique radiative Higgs corrections and they can be omitted. We shall concentrate on the oblique radiative corrections involving vacuum polarization. We see that for $m_H=125$ GeV the Higgs contribution (\ref{Higgsdelta}) becomes negligible and from (\ref{Higgsdelta}), we obtain $\Delta\rho_{H(1)}\sim 10^{-4}$.

The radiative corrections to EW observables in the SM are parameterized, so that any contributions to new physics beyond the standard model are implemented and compared with experimental data. We assume that the group is still $SU(3)_c\times SU(2)_L\times U(1)_Y$ and that it couples only to the observed standard model particles: the 12 quarks and leptons, the $W^{\pm}$, the $Z^0$, the Higgs and the photon. Neglecting all direct box and vertex corrections, we consider only the oblique corrections affecting the $\gamma, Z, W$ two-point functions and the $Z\gamma$ mixing. The standard oblique $S,T,U$ parametrization of physics beyond the standard model~\cite{Peskin2,Altarelli}, can predict new heavy fermions in the vacuum polarization radiative corrections. Our model does not have any new heavy fermions beyond the observed 12 quarks and leptons, nor does our minimal model have new $Z'$ or $W'$ gauge bosons.  However, we have to consider the sensitivity of our model to oblique Higgs loop corrections. The prediction of a certain EW observable $O$ is given by the sum:
\begin{equation}
O=O_{\rm M,ref}(m_H,m_t)+c_sS+c_TT+c_UU.
\end{equation}
The $S,T,U$ parameters measure deviations from the EW model, $M_{\rm ref}$, and vanish if the data are equal to the $M_{\rm ref}$ prediction.

The energy scales at which we consider the effects of the $S,T,U$ parameters is set by $\Lambda_M$, so the vacuum polarization functions are expanded in powers of $p^2/\Lambda_M^2$ and $p^2/\Lambda_H^2$, keeping only the constant and linear terms in $p^2$. The constant pieces $\Pi_{\gamma\gamma}(0)$ and $\Pi_{Z\gamma}(0)$ are zero because of the Ward identities for the gauge invariant $U_{\rm em}$ interactions. The standard $S,T,U$ parameters are chosen to be~\cite{Peskin2,Altarelli}:
\begin{align}
\alpha S&=4\sin^2\theta_w\cos^2\theta_w\biggl[\Pi'_{ZZ}(0) -\frac{\cos^2\theta_w-\sin^2\theta_w}{\sin\theta_w\cos\theta_w}\Pi'_{Z\gamma}(0)-\Pi'_{\gamma\gamma}(0)\biggr],\nonumber\\
\alpha T&=\frac{\Pi_{WW}(0)}{M_W^2}-\frac{\Pi_{ZZ}(0)}{M_Z^2},\nonumber\\
\alpha U&=4\sin^2\theta_w\biggl[\Pi'_{WW}(0)-\cos^2\theta_w\Pi'_{ZZ}(0)-2\sin\theta_w\cos\theta_w\Pi'_{Z\gamma}(0)
-\sin^2\theta_w\Pi'_{\gamma\gamma}(0)\biggr],
\end{align}
where $\Pi'$ denotes the derivative of the vacuum polarization $\Pi$ with respect to $p^2$.

The Higgs boson action term $HD_\mu H^2/\Lambda_H^2$ (where $D_\mu$ is the covariant derivative) only contributes to $T$ and $U$. The action term $HW^{\mu\nu}B_{\mu\nu}H/\Lambda_H^2$ contributes to $S$ and not to $T$ or $U$, while $(HW^{\mu\nu} H)(HW_{\mu\nu}H)/\Lambda_M^4$ contributes to $U$.

The $S$ parameter determines Higgs contributions to neutral currents, while $T$ measures the difference between neutral and charged Higgs corrections, and affects the measurements of the parameter $\rho$. The $U$ parameter makes small corrections to the model.

\section{Higgs Mass Hierarchy and Fine Tuning Problem and the Stability of the Vacuum}

Let us investigate the perturbative treatment of the Higgs boson mass and the radiative corrections. The Lagrangian we consider for a real scalar field describing the Higgs boson is given by
\be
{\cal L}_H=\frac{1}{2}\phi\Box\phi-\frac{1}{2}m_0^2\phi^2-\frac{1}{4!}\lambda_0\phi^4.
\ee
The field strength renormalization constant $Z$ and the bare parameters $m_0$ and $\lambda_0$ are series expansions in powers of the coupling $\lambda$, the physical mass $m_H$ and the energy scale $\Lambda_H$:
\be
Z=1+\delta Z(\lambda,m_H^2,\Lambda_H^2),
\ee
\be
Zm_0^2=m^2+\delta m_H^2(\lambda,m_H^2,\Lambda_H^2),
\ee
\be
Z^2\lambda_0=\lambda+\delta Z(\lambda,m_H^2,\Lambda_H^2).
\ee

The entire function distribution is given by
\be
{\cal E}=\exp\biggl(\frac{\Box+m_H^2}{2\Lambda_H^2}\biggr).
\ee
The propagator in Euclidean momentum space using the Schwinger proper time formalism is
\be
i\tilde\Delta_H(p)\equiv \frac{i{\cal E}^2}{p^2+m_H^2}=i\int\frac{d\tau}{\Lambda_H^2}
\exp\biggl[-\tau\biggl(\frac{p^2+m_H^2}{\Lambda_H^2}\biggr)\biggr].
\ee

The one-loop Higgs boson self-energy graph can be evaluated to give the result:
\be
\label{OneLoop}
-i\delta\Sigma=\frac{-iZ^{-2}\lambda}{32\pi^2}m_H^2\Gamma\biggl(-1,\frac{m_H^2}{\Lambda_H^2}\biggr),
\ee
where $\Gamma(n,z)$ is the incomplete gamma function:
\be
\label{Gamma}
\Gamma(n,z)=\int_z^\infty dtt^{n-1}\exp(-t)=(n-1)\Gamma(n-1,z)+z^{n-1}\exp(-z).
\ee
An asymptotic expansion in $\Lambda_H$ can be obtained by using the recursion relation (\ref{Gamma}) to reach $\Gamma(0,z)=E_1(z)$ where $E_1$ is the exponential integral:
\be
E_1(z)=\int_z^\infty dt\frac{\exp(-t)}{t}=-\ln(z)-\gamma-\Sigma^{\infty}_{n=1}\frac{(-z)^n}{nn!}.\ee

We can write (\ref{OneLoop}) in the form:
\be
\Sigma(p^2)=\delta Z(p^2-m_H^2)+\delta m_H^2
+\frac{Z^{-1}\lambda}{32\pi^2}m_H^2\Gamma\biggl(-1,\frac{m_H^2}{\Lambda_H^2}\biggr)+{\cal O}(\lambda^2).
\ee
From these considerations, we obtain the mass and field strength finite renormalizations~\cite{KleppeWoodard1993}:
\be
\delta m_H^2=-\frac{\lambda}{32\pi^2}m_H^2\Gamma\biggl(-1,\frac{m_H^2}{\Lambda_H^2}\biggr)+{\cal O}(\lambda^2),
\ee
and
\be
\delta Z={\cal O}(\lambda^2).
\ee
The expansion of the one-loop Higgs boson mass correction is
\be
\label{masscorrection}
\delta m_H^2=\frac{\lambda}{32\pi^2}\biggl[-\Lambda_H^2+m_H^2\ln\biggl(\frac{\Lambda_H^2}{m_H^2}\biggr)
+m_H^2(1-\gamma)+{\cal O}\biggl(\frac{m_H^2}{\Lambda_H^2}\biggr)\biggr]+{\cal O}(\lambda^3).
\ee
The one-loop vertex correction is obtained as
\be
\label{vertexcorrection}
\delta\lambda=\frac{3\lambda^2}{16\pi^2}\int_0^{1/2}dx\Gamma\biggl(0,\frac{1}{1-x}\frac{m_H^2}{\Lambda_H^2}\biggr)
+{\cal O}(\lambda^2).
\ee
This can be expanded to give
\be
\delta\lambda=\frac{3\lambda^2}{16\pi^2}\biggl[\frac{1}{2}\ln\biggl(\frac{\Lambda_H^2}{m_H^2}\biggr)
+\frac{1}{2}(\ln(2)-1-\gamma)+{\rm O}\biggl(\frac{m_H^2}{\Lambda_H^2}\biggr)\biggr]+{\rm O}(\lambda^3).
\ee
We now choose $\Lambda_H\sim 200\,{\rm GeV}-1\,{\rm TeV}$ as the Higgs boson entire function distribution energy scale. The mass radiative corrections in (\ref{masscorrection}) are reduced leading to $\delta m_H^2/m_H^2\sim {\cal O}(1)$.

A resolution of the Higgs boson mass and fine-tuning problem can now be obtained. The dominant Higgs particle loop contribution is
\be
\delta m_H^2=\frac{y_t^2}{16\pi^2}\Lambda_H^2+\delta{\cal O}(m_{\rm weak}^2).
\ee
With the choice $\Lambda_H\sim 200\,{\rm GeV}-1\,{\rm TeV}$, we have $\delta m_H^2/m_H^2\sim {\cal O}(1)$ and the Higgs-loop radiative corrections are damped for $p^2 > \Lambda_H^2$.

A stable, true vacuum is determined by the global minimum of the scalar potential, which depends on the particle model's scalar fields. The spontaneous symmetry breaking of the vacuum in the SM produces a ``false'' vacuum, instead of a local minimum. If only the deepest minimum of the scalar field potential is occupied by the Universe, then its future is not threatened by an unstable vacuum. A local minimum in the current Universe, can become a deeper minimum in the potential, and only a finite barrier separates the potential from a bottomless pit, and the Universe can tunnel out into another state in which it cannot support life as we know it~\cite{Kusenko}. The SM potential (\ref{GinsbergLandaupotential}) allows for a renormalizable QFT in the SM with the scalar field $\phi$ restricted to the fourth power. It has a false vacuum whose evolution depends on the measured Higgs mass $M_H=125$ GeV and the coupling constant $\lambda$. The nature of the vacuum and the existence of the true vacuum can be calculated by using renormalization group evolution~\cite{Bednyakov}). A predicted instability (or metastability) implies that the SM cannot be valid all the way to the Planck energy scale, suggesting that new particles must contribute to the scalar field potential. Understanding the role of the scalar field potential can shed light on the cause of the matter-antmatter asymmetry.

In our finite QFT particle model, we do not require infinite renormalizability, nor an initial massless $SU(2)\times U(1)$. We do not invoke a spontaneously broken vacuum state based on the scalar field potential (\ref{GinsbergLandaupotential}) with an imaginary mass and a false vacuum. Consequently, we can expect that the scalar field potential for the Universe lies in the deepest minimum of the potential with $\langle 0|\phi_H|0\rangle=0$ and there is no threat to the stability of the vacuum and the future evolution of the Universe.

\section{Experimental Tests of the Model}

Future Higgs boson experimental data, which will determine the sizes of Higgs loop radiative corrections, can distinguish our finite QFT model from the SM. This can check whether radiative Higgs-fermion and Higgs self-energy loops can experimentally satisfy the restriction $\Lambda_H\lesssim 1\,{\rm TeV}$ on their magnitudes. Future accelerator experiments at energies greater than $\Lambda_M > 1-14\,{\rm TeV}$ can test whether scattering amplitudes and cross sections are damped by the entire function operator ${\cal E}(p^2)$ for $\Lambda_M > 1-14$ TeV compared to the predicted scattering amplitudes and cross sections of the SM.

If no new fundamental particles are detected by the LHC and future high energy accelerators, what explains the lack of new fundamental particles up to the Planck energy, $E\sim 10^{19}\,{\rm GeV}$? Because the scattering amplitudes predicted by the finite QFT model are damped exponentially fast in Euclidean momentum space for $E > \Lambda_M $, the ``desert" energy hierarchy problem between the EW energy scale $\sim 246$ GeV and the Planck energy scale $\sim 10^{19}$ GeV can be explained.

The Higgs particle is described by a resonance and the processes we are concerned with are far from the resonance peak at $M_H=125$ GeV. The same is true for the $W^\pm$ and $Z^0$ resonances. All the data for the Higgs boson interactions obtained from the LHC scattering of protons is determined by the Higgs boson decay products. Let us consider the self-energies. The imaginary part of a self-energy diagram $A_S$ with internal lines of masses $m_1$ and $m_2$ is given by~\cite{Anselmi}:
\be
{\rm Im}(A_S)\propto \theta[s-(m_1+m_2)^2]\biggl(1-\frac{(m_1+m_2)^2}{s}\biggr)^{1/2}\biggl(1-\frac{(m_1-m_2)^2}{s}\biggr)^{1/2},
\ee
where $s$ is the center-of-mass energy squared. Denote by ${\cal I}^\gamma_{W^+W^-}$, ${\cal I}^Z_{W^+W^-}$ and ${\cal I}^Z_{ZH}$ the imaginary parts of the respective bubble graphs. Measurements can determine their contributions depending on their energy thresholds. These ${\cal I}$ can be measured by the cross section $\sigma(e^+e^-\rightarrow {\rm leptons,hadrons})$ and are in the range of the LEP accelerator, which had a maximum electron-positron collision energy of 200 GeV. No damping of the cross sections for ${\cal I}^\gamma_{W^+W^-}$, ${\cal I}^Z_{W^+W^-}$ was detected. The same was true for the leptons and the SM quarks, except for the top quark due to its high energy threshold. The case of the Higgs boson remains unresolved, for ${\cal I}^\gamma_H, {\cal I}^Z_H$ and ${\cal I}^{\gamma Z}_H$ with circulating top quarks are undetected due to their energy thresholds $2m_t\sim 344$ GeV. However, precision measurements at a future International Linear Collider (ILC) or at a future Circular Electron-Positron Collider (CEPC) could measure the Higgs boson self-energy contributions. The LEP has measured the box diagram with a threshold equal to $m_Z$ and no visible significant damping of the self-energy amplitudes was detected.  We can deduce that the $\gamma$ and $Z$ particles are at energies $E < \Lambda_M$. The future collider experiments can measure the physical Higgs boson ${\cal I}_H$, and if it is found that the Higgs boson couplings and self-energies making up the radiative loop corrections are significantly damped at energies above $\sqrt{s} > 250$ GeV, then we are justified to choose $\Lambda_H\lesssim 1$ TeV, thereby, solving the Higgs mass hierarchy problem.

\section{Conclusions}

We have formulated an alternative to the SM model based on a finite Poincar\'e invariant and unitary QFT. The model accepts that the EW $SU(2)\times U(1)$ group is broken by the $W^\pm$, $Z^0$, the quark and lepton masses and the Higgs mass. The finite QFT avoids the need for an infinite renormalization of loop diagrams. The loop diagrams are made finite by attaching infinite derivative, entire function distribution operators to loop vertices and propagators, which are exponentially damped in Euclidean momentum space. The rate of the momentum space damping is controlled by the physical energy scales $\Lambda_M=1/\ell_M$ and
$\Lambda_H=1/\ell_H$. The loop diagrams for $W^\pm$, $Z^0$ and fermion couplings are controlled by the energy scale $\Lambda_M$. All the experimental data obtained at the CERN LHC will be satisfied by the alternative finite QFT model for scattering amplitudes and particle decay rates with $E < \Lambda_M$. By choosing the Higgs self-coupling energy scale, $\Lambda_H\lesssim  1\,{\rm TeV}$, we resolve the Higgs mass fine-tuning and naturalness problem.

The damping of scattering amplitudes and cross sections for $p^2 >\Lambda^2_M$ and the damping of Higgs boson loops for $p^2 >\Lambda^2_H$ guarantee that {\it the finite QFT is asymptotically safe}. These asymptotically safe predictions of the model can be tested in future High Energy linear and Circular Accelerators.

We have demonstrated that the nonlocal nature of the field operators due to entire function distribution operators at vertices and in propagators {\it does not lead to violations of microcausality}, by extending the lack of causality violation in non-relativistic quantum mechanics to relativistic QFT.

The finite QFT has been extended to quantum gravity~\cite{Moffat1990,MoffatWoodard1991,Moffat2011,Moffat2014,Biswas2005,Biswas2012}. The perturbative quantum gravity theory is formulated as an expansion around a fixed background metric ${\bar g}_{\mu\nu}$:
\be
g_{\mu\nu}={\bar g}_{\mu\nu}+h_{\mu\nu},
\ee
where $h_{\mu\nu} < 1$ and ${\bar g}_{\mu\nu}$ can be chosen to be the Minkowski metric $\eta_{\mu\nu}$.

The quantum gravity theory is finite and unitary to all orders of perturbation theory and is based on the Einstein-Hilbert Lagrangian linear in the curvature. A consequence of the finite quantum gravity theory is that graviton scattering amplitudes, including graviton loops coupled to gravitons and matter, are damped in Euclidean momentum space and quantum gravity is asymptotically safe. A solution of the cosmological constant problem can be achieved using the finite quantum gravity theory~\cite{Moffat2011(2),Moffat2014}.

\section*{Acknowledgments}

I thank Latham Boyle, Emil Mottola, Mark Wise, Martin Green and Viktor Toth for helpful discussions. This research was supported in part by Perimeter Institute for Theoretical Physics. Research at Perimeter Institute is supported by the Government of Canada through the Department of Innovation, Science and Economic Development Canada and by the Province of Ontario through the Ministry of Research, Innovation and Science.

\end{document}